\documentclass[bibyear]{aa}  
\usepackage{graphicx}
\usepackage{txfonts}
\usepackage[normalem]{ulem}
\usepackage{color}

\begin{document}

   \title{THA~15$-$31: Discovery with VLT/X-Shooter and Swift/UVOT 
          of a new symbiotic star of the accreting-only variety}
   \author{U. Munari
          \inst{1}
          \and
          J.~K. Alcal{\'a}
          \inst{2}
          \and
          A. Frasca
          \inst{3}
          \and
          N. Masetti
          \inst{4,5}
          \and
          G. Traven
          \inst{6}
          \and
          S. Akras
          \inst{7}
          \and
          L. Zampieri
          \inst{1}
          }
   \institute{  INAF-Osservatorio Astronomico di Padova, 36012 Asiago (VI), Italy,\\
              \email{ulisse.munari@inaf.it}
         \and
                INAF-Osservatorio Astronomico di Capodimonte, via Moiariello 16, 80131 Napoli, Italy
         \and
                INAF-Osservatorio Astrofisico di Catania, via S. Sofia 78, 95123 Catania, Italy
         \and
                INAF-Osservatorio di Astrofisica e Scienza dello Spazio, via Gobetti 101, 40129 Bologna, Italy           
         \and
                Departamento de Ciencias F\'isicas, Universidad Andr\'es Bello, Fern\'andez Concha 700, Las Condes, Santiago, Chile
         \and
                Lund Observatory, Department of Astronomy and Theoretical Physics, Box 43, SE-221 00 Lund, Sweden
         \and
                Institute for Astronomy, Astrophysics, Space Applications and Remote Sensing, National Observatory of Athens, Penteli GR 15236, Greece 
             }

   \titlerunning{Discovery of a new symbiotic star}   

   \date{Received ; accepted }

  \abstract{We report the discovery and characterization of a new symbiotic
 star of the accreting-only variety, which we observed in the
 optical/near-infrared (NIR) with VLT/X-Shooter and in the
 X-rays/ultraviolet with Swift/UVOT+XRT.  The new symbiotic star,
 THA~15$-$31, was previously described as a pre-main sequence star belonging
 to the Lupus~3 association.  Our observations, ancillary data, and Gaia
 EDR3 parallax indicate that THA~15$-$31 is a symbiotic star composed of an
 M6III red giant and an accreting companion, is subject to $E_{B-V}$=0.38
 reddening, and is located at a distance of $\sim$12 kpc and at 1.8 kpc
 above the Galactic plane in the outskirts of the Bulge.  The luminosity of
 the accreting companion is $\sim$100~L$_\odot$, placing THA~15$-$31 among
 the symbiotic stars accreting at a high rate
 (2.5$\times$10$^{-8}$~M$_\odot$\,yr$^{-1}$ if the accretion is occurring on
 a white dwarf of 1~M$_\odot$).  The observed emission lines originate
 primarily from HI, HeI, and FeII, with no HeII or other high-excitation
 lines observed; a sharp central absorption superimposed on the Balmer
 emission lines is observed, while all other lines have a simple
 Gaussian-like profile.  The emission from the companion dominates over the
 M6III red giant at $U$ and $B$-band wavelengths, and is consistent with an
 origin primarily in an optically thick accretion disk.  No significant
 photometric variability is observed at optical or NIR wavelengths,
 suggesting either a face-on orbital orientation and/or that the red giant
 is far from Roche-lobe filling conditions.  The profile of emission lines
 supports a low orbital inclination if they form primarily in the accretion
 disk.  An excess emission is present in AllWISE W3 (12$\mu$m) and W4
 (22$\mu$m) data, radiating a luminosity $\geq$35~L$_\odot$, consistent with
 thermal emission from optically thin circumstellar dust.}

 \maketitle

\section{Introduction}

THA~15$-$31 was first noted by The (1962) while compiling a list of faint
H-alpha emission stars in Lupus and Scorpius, and later rediscovered as
Sz~105 by Schwartz (1977) during a deep red objective-prism survey of
southern dark clouds.  This led to THA~15$-$31 being considered as a
candidate young stellar object (YSO) and as such it was included in many
surveys of the Lupus~3 star forming region, with observations ranging from
X-rays to optical, infrared (IR), and radio (Hughes et al.  1994; Chen et
al.  1997; Ghez et al.  1997; Krautter et al.  1997; Nuernberger, Chini, \&
Zinnecker 1997; G{\'o}mez \& Persi 2002; Kirk \& Myers 2011; Mu{\v{z}}i{\'c}
et al.  2014).  The classification of THA~15$-$31 as a classical T~Tau star
of M4 spectral type with 0.15~M$_\odot$ mass and 0.03~L$_\odot$ luminosity
was generally copied from one survey to another.

The first hint of an erroneous YSO classification of THA~15$-$31 came from
L{\'o}pez Mart{\'\i} et al.  (2011) who investigated the distribution in
proper motion of members of Lupus~3 and found that THA~15$-$31 belongs to a
heterogeneous group of background objects that are highly distinct from true
YSOs in the foreground.  The problematic absence of LiI 6708 absorption in
the spectra of THA~15$-$31 ---a hallmark of YSOs--- was then noted by
Alcal{\'a} et al.  (2014), while Frasca et al.  (2017) derived a surface
gravity for THA~15$-$31 that is more in line with a cool giant than a YSO. 
Based on these findings, THA~15$-$31 was then finally rejected by Alcal{\'a}
et al.  (2017) from the list of YSO members of Lupus~3.  This is, in broad
terms, the current and rather limited knowledge about THA~15$-$31.

Our interest in THA~15$-$31 was triggered by the recent release of Gaia EDR3
(Gaia Collaboration 2020), which lists a 0.0422$\pm$0.0398 mas parallax for
the object.  Although still affected by a large error, it significantly
improves upon the previous value of 0.0323$\pm$0.0765 mas given in Gaia~DR2,
and definitively supports a large distance (at least several kpc) to
THA~15$-$31, which in turn implies the presence of a cool giant in the
system and definitively rules out any link to the Lupus dark clouds.  To
investigate its nature, we analyzed an archive VLT/X-Shooter optical-IR
spectrum of THA~15$-$31 and obtained satellite ultraviolet (UV) observations
with the Neil Gehrels Swift Observatory (hereafter Swift).  Supplemented by
epoch photometry from NeoWISE and ASASSN, as well as other ancillary data,
we now show that THA~15$-$31 is indeed an accreting-only symbiotic star
(SySt), with an accretion luminosity ($L_{\rm acc}$) that is much higher
than that characterizing the sample of 33 new, accreting-only SySts that we
recently identified and studied (Munari et al.  2021) among the 0.6 million
stars observed by the GALAH spectroscopic survey of the southern sky (de
Silva et al.  2015, Buder et al.  2021).

Symbiotic stars are interacting binaries where a red giant (RG) fuels a hot
companion (HC) via accretion either through Roche-lobe overflow or wind
intercept.  In the vast majority of SySts the hot companion is believed to
be a white dwarf (WD), while the hard and pulsed X-ray emission observed in
a small subset suggests that the accretion takes place on a neutron star
(NS).  The RG+WD symbiotic stars are broadly divided into two major groups
(see the recent review by Munari 2019 for details): those {\it
accreting-only} ({\it acc}-SySts) whose optical spectra are dominated by the
RG with no or weak emission lines, and the {\it burning-type} ({\it
burn}-SySts) with optical spectra dominated by a strong nebular continuum
and a rich emission line spectrum, that originate from the wind of the RG
which is largely ionized by the very hot and luminous WD undergoing surface
nuclear burning of accreted material.

It is possible that SySts spend most of their time in the accreting-only
phase, quietly accumulating material on the surface of the WD.  When enough
material has built up, nuclear burning begins.  If the accreted matter is
electron degenerate, the burning proceeds explosively (Starrfield et
al.~2020) resulting in a nova outburst.  If the accreted matter is instead
not electron-degenerate, the nuclear burning develops slowly and in thermal
equilibrium (Fujimoto et al.~1982).

There is a clear disproportion among cataloged SySts (Allen 1984,
Belczy{\'n}ski et al.  2000, Merc et al.  2019, Akras et al.  2019a) in
favor of the {\it burn}-SySt type which are much more easily discovered, and
the known examples of the {\it acc}-SySt variety could simply be ``the tip
of the iceberg'' (Mukai et al.~2016).  The {\it acc}-SySts are well
distributed in accretion luminosity, from the 10$-$100 L$_\odot$ of some
well-studied objects like EG And (Skopal 2005, Nu{\~n}ez et al.~2016) to the
emerging large population of lower 1$-$10 L$_\odot$ objects exemplified by
the first sample of 33 SySts discovered by Munari et al.  (2021) in the
course of the GALAH survey.  The higher $L_{\rm acc}$ systems are
characterized by a brighter HC (collectively including the accretion disk,
its central star, and some RG wind ionized by them both) and by much
stronger emission lines compared to lower $L_{\rm acc}$ systems, although
generally missing the higher ionization features like HeII, [FeVII], OVI, or
[FeX] that adorn the spectra of the {\it burn}-SySt type.  SySts accreting
on neutron stars, while generally inconspicuous at optical and near-UV
wavelengths, may radiate abundantly in hard X-rays (e.g., Masetti et al. 
2007a,b, Nucita et al.  2014).

Identification of new SySts of the accreting-only, high-$L_{\rm acc}$
variety appears to be important for properly measuring their frequency
relative to the lower $L_{\rm acc}$ systems, and will help to address the
fundamental question of whether accretion luminosity can be used to
distinguish between SySts of different nature (e.g., in terms of mass,
orbital separation, and filling of the Roche lobe), or whether accretion
simply proceeds in a highly sporadic fashion, irrespective of the nature of
the underlying SySt.  The latter applies to the symbiotic star and recurrent
nova T~CrB, as discussed in Munari et al.  (2016) and Luna et al.  (2020):
for 70 years following its nova eruption in 1946, the accretion luminosity
was so low that only feeble H$\alpha$ emission was generally visible
superimposed on the spectrum of the M3III cool giant.  However, starting
with 2015, the $L_{\rm acc}$ in T~CrB hugely increased and the optical
spectra are now persistently dominated by a strong nebular continuum and
imposing emission lines, in a replica of the accretion surge that preceded
the 1946 thermonuclear outburst (Hachenberg \& Wellmann 1939).

   \begin{figure*}
   \centering
   \includegraphics[angle=270,width=17.5cm]{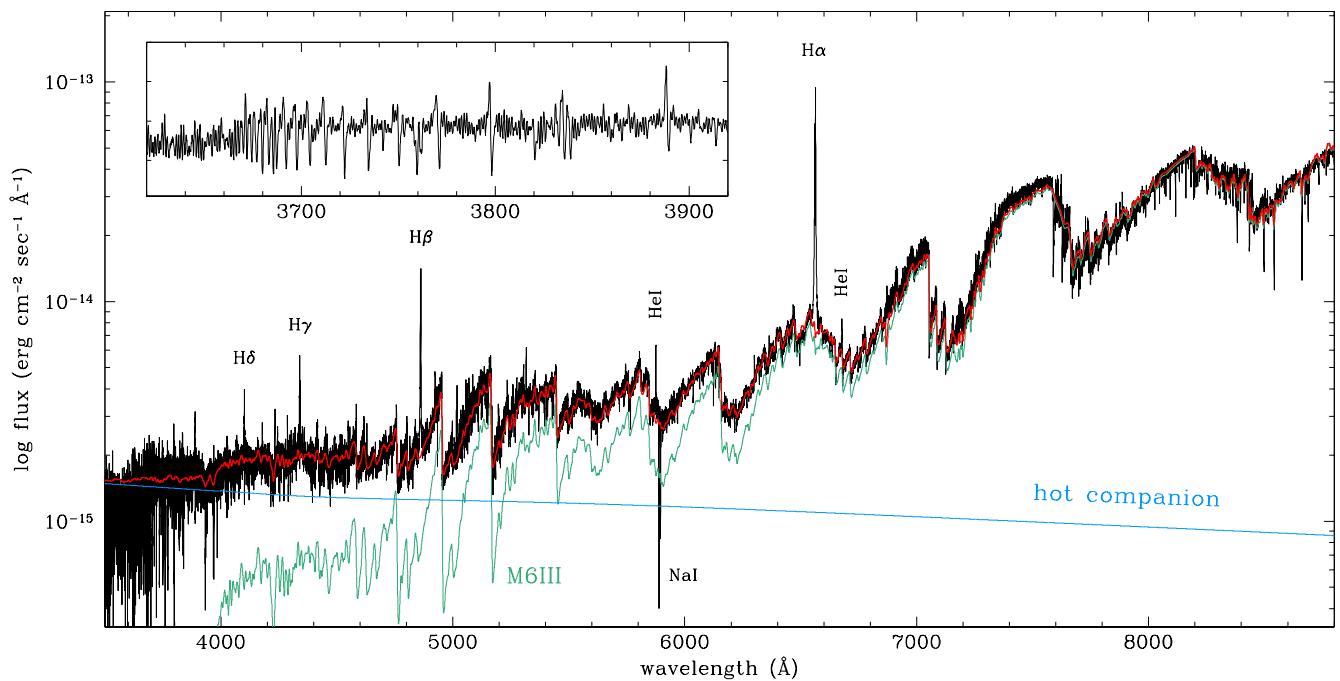}
      \caption{Optical spectrum of THA~15$-$31 as recorded with X-Shooter
      (black) is fitted (red) by combining the spectrum of the cool giant
      (M6III, green spectrum) and that of the hot companion (blue line). 
      The latter is a bremsstrahlung spectrum taken to approximate the mix
      of the accretion disk, the central star, and possibly some RG wind
      ionized by them.  The spectra of the M6III cool giant and of the hot
      companion are reddened by $E_{B-V}$=0.38 following the Fitzpatrick
      (1999) reddening law for the standard $R_V=3.1$ case.  The inset is a
      zoomed-in view of the head of the Balmer series and the Balmer jump.}
         \label{fig:M6III}
   \end{figure*}

\section{Observations}

\subsection{X-Shooter spectrum}

The X-Shooter data used here were acquired on April 18, 2012, within the
context of the X-Shooter guaranteed-time observations (GTO) granted to INAF
(cf.  Alcal{\'a} et al.  2011, 2014 and references therein).  X-Shooter
provides simultaneous wavelength coverage from $\sim$3000~\AA\ to $\sim$2.48
$\mu$m.  For the observations of THA\,15-31, slits of
0\farcs5/0\farcs4/0\farcs4 were used in the 
UltraViolet-Blue/Visual/Near-InfraRed (UVB/VIS/NIR) arms, respectively,
yielding resolving powers of 9100/17400/10500.

The observations were performed at parallactic angle with one cycle using
the A-B nodding mode, with single-node exposure of 150\,s, 100\,s, and
100\,s for a total of 300\,s, 200\,s, and 200\,s in the UVB, VIS, and
NIR arms, respectively.  Telluric standard stars were observed with the same
instrumental setup and at very similar airmass to THA\,15-31.  Flux
standards during the same night were observed through the 5\,arcsec slits
for flux calibration purposes.

Data reduction was performed independently for each spectrograph arm using
the X-Shooter pipeline (version 1.3.7; Modigliani et al.  2010).  The
standard steps of processing included bias and dark subtraction,
flat-fielding, optimal extraction, sky subtraction, and wavelength and flux
calibration (the latter carried out in IRAF\footnote{IRAF is distributed by
the National Optical Astronomy Observatories, which are operated by the
Association of Universities for Research in As- tronomy, Inc., under
cooperative agreement with the National Science Foun- dation.}).  Wavelength
shifts due to instrumental flexure were corrected for using the {\sc
flexcomp} package within the pipeline.  The precision in wavelength
calibration is better than 0.01\,pix in the UVB and VIS arms, corresponding
to 0.02\,\AA, but errors can be as large as $\sim$0.06\,\AA\ in the NIR arm. 
The wavelength calibration allows an average precision in radial velocity of
$\sim$1.0\,km/s in the VIS arm.  The telluric correction was performed
independently in the VIS and NIR spectra, as explained in Appendix~A of
Alcal{\'a} et al.  (2014).  By comparing the response function of different
flux standards observed during the same night, we estimate an intrinsic
precision on the flux calibration of $\sim$5\% (cf.  Alcal{\'a} et al. 
2014, 2017).

\begin{table}[t]
\small
\begin{center}
\caption{Swift-UVOT observations of THA~15$-$31.}
\label{swift}
\begin{tabular}{cccc}
\hline\hline
filter& $\lambda_c$ & F$_\lambda$ & percent         \\ 
      & (\AA)       & ($\times$10$^{-16}$ erg cm$^{-2}$ s$^{-1}$ \AA$^{-1}$) & error \\
\hline
UVW1  &  2600 &  2.9  & 10\%  \\            
UVM2  &  2246 &  1.1  & 20\%  \\           
UVW2  &  1928 &  1.5  & 14\%  \\           
\hline
\end{tabular} 
\end{center}
\end{table}

\subsection{Swift observations}

Our {\it Swift} observations of THA~15$-$31 were acquired on April 9, 2021,
starting at 21:43 UT with the on-board instruments X-Ray Telescope
(hereafter XRT, Burrows et al.  2005) and UltraViolet-Optical Telescope
(hereafter UVOT, Roming et al.  2005).  The XRT allows coverage of the X--ray
band between 0.3 and 10 keV, whereas UVOT data were collected using the
three available UV filters (cf.  Poole et al.  2008 and Breeveld et al. 
2011 for details).  On-source pointing was simultaneously performed with the
two instruments; it lasted 1560~s for XRT, whereas for UVOT 444 s were spent
using the UVW1 filter and 515 s in both UVM2 and UVW2 filters.

Count rates on Level 2 (i.e.  calibrated and containing astrometric
information) UVOT images at the position of our target were measured through
aperture photometry using 5$''$ apertures, whereas the corresponding
background was evaluated using a combination of several circular regions in
source-free nearby areas.  Magnitudes were measured with the {\sc
uvotsource} task.  The data were then calibrated using the UVOT photometric
system described by Poole et al.  (2008), and we included the recent
(November 2020) fixings recommended by the UVOT team.  The observed fluxes
in the three UV filters are listed in Table~\ref{swift} and the
corresponding magnitudes in Table~\ref{summary} (Vega scale).

THA~15$-$31 was not detected in X-rays with XRT.  Data analysis was
performed using the {\sc xrtdas} standard pipeline package ({\sc
xrtpipeline} v.  0.13.4) in order to produce screened event files.  All
X-ray data were acquired in photon counting (PC) mode (Hill et al.  2004)
adopting the standard grade filtering (0-12 for PC) according to the XRT
nomenclature.  For each source, scientific data were extracted from the
images using an extraction radius of 47$''$ (20 pixels) centered at the
optical coordinates of the source, while the corresponding background was
evaluated in a source-free region of radius 94$''$ (40 pixels).

We do not detect any X-ray emission down to a 3$\sigma$ limit of
2.2$\times$10$^{-3}$ cts s$^{-1}$ in the 0.3--10 keV band at the position of
THA~15$-$31; this estimate was obtained using the {\sc xspec} package. 
Assuming a bremsstrahlung spectrum, with temperature $kT$ = 2 keV, which
gives a 0.3--10 keV flux limit of $<$6.3$\times$10$^{-14}$ erg cm$^{-2}$
s$^{-1}$.

\section{Results}

\subsection{Spectral type, radial velocity, and reddening}

The optical part of the X-Shooter spectrum shown in Figure~\ref{fig:M6III}
is dominated by the molecular TiO absorption bands of the RG, and this
offers the possibility to derive some of the basic properties of
THA~15$-$31: (1) the relative intensity of molecular bands and their fine
structure, which vary by huge amounts among cool giants, thus allowing us to
derive an accurate spectral type for the RG; (2) the veiling by the emission
from HC, which increases toward shorter wavelengths; and (3) the overall
slope, controlled by the amount of reddening.  The reddening and spectral
type are determined without significant degeneracy among them, contrary to
the case of modeling the spectral energy distribution (SED) based on
broad-band photometric data only: the key to success here is the the strong
and sharp band-heads that develop within individual TiO bands over a short
wavelength interval to make their interplay independent from reddening.  The
technique was thoroughly tested and applied in Corradi et al.  (2010) for
the analysis of new SySts discovered photometrically by the INT Photometric
H$\alpha$ Survey (IPHAS) and re-observed spectroscopically for confirmation
and characterization.

The fitting to the optical part of the X-Shooter spectrum of THA~15$-$31 is
presented in Figure~\ref{fig:M6III}.  To this aim we adopted the
library of intrinsic spectra of real M-type giants by Fluks et al. 
(1994)\footnote{We corrected an incorrect shape in the Fluks et al. 
spectrum over the 7300$-$8200~\AA\ range and replaced the linear
extrapolation below 3900~\AA\ by substituting them with observations of HD
148783 that we obtained with the Asiago 1.22m+B\&C telescope.  This star is
the prime standard for spectral type M6III according to the compilation by
Yamashita et al. (1977).} and the interstellar reddening law of
Fitzpatrick (1999) for the standard $R_V$=3.1 case.  The emission from HC is
a mixture of different sources (accretion disk, central star, nebular
continuum) present in unknown proportions, and physical parameters that
cannot be unambiguously determined with available data.  As a fair
approximation of the overall shape of the HC emission in the wavelength
range covered by Figure~\ref{fig:M6III}, we adopted a bremsstrahlung
spectrum for gas at $T_e$=10,000~K.  However, none of the results presented
in this paper depend on this assumption.

The fit of the X-Shooter spectrum displayed in Figure~\ref{fig:M6III} is
remarkably good, fixing the reddening to $E_{B-V}$=0.38$\pm$0.04 and the
spectral type to M6III, with an uncertainty smaller than 1 spectral
subclass, because the structure of molecular bands is strongly different for
stars with M5III and M7III spectral types.  A spectral type M6III is the one
most frequently found in Galactic SySts.  The modest $E_{B-V}$=0.38$\pm$0.04
reddening agrees well with the relatively high Galactic latitude of the
object ($b$=+8.5 deg), and with the {\sc Stilism} 3D-map of the Galactic
interstellar reddening (Lallement et al.  2014, Capitanio et al.  2017): the
latter indicates an exit of the line of sight to THA~15$-$31 from the dust
slab on the Galactic plane at a distance of 600~pc from Earth, where it
reaches $E_{B-V}$=0.40$\pm$0.05.  The agreement is also good with the value
$E_{B-V}$=0.34$\pm$0.03 derived from the 3D map of Galactic extinction by
Marshall et al.  (2006).  Unfortunately, the location on the sky of
THA~15$-$31 is too far south for other popular 3D maps of Galactic
extinction to be considered, such as those based on IPHAS (Sale et al. 
2014) or PanSTARRS (Green et al.  2019) sky surveys.  The limited resolution
of X-Shooter is not sufficient to attempt to resolve the interstellar from
the stellar component of the NaI line, and thus derive the reddening from
its equivalent width (see also below, Sect.~3.6).
The heliocentric radial velocity of the M6III giant is determined as
\begin{equation}
{\rm RV}_\odot = +4 \pm 2 {\rm km~s^{-1}}
\end{equation}
by application of the {\sc rotfit} code (Frasca et al.  2006), specifically
modified for X-Shooter data (cf Alcal{\'a} et al.  2014), and confirmed by
cross-correlation against the synthetic spectral library of Munari et al. 
(2005).

The inset in Figure~\ref{fig:M6III} zooms onto the THA~15$-$31 spectrum at
the head of the Balmer series, highlighting two interesting features that
are further discussed in Sect.~4.  First, the series of individual Balmer
lines is visible in absorption up to about H(29), corresponding to much
higher quantum levels than those observed in normal stars.  Second, the
Balmer continuum below $\lambda \leq 3646$~\AA\ is in absorption.  This is a
rare occurrence: inspection of the spectral atlases of SySt by Allen (1984)
and Munari and Zwitter (2002), which survey most ($\sim$150) of the
symbiotic stars known at the time, reveals only about half a dozen objects
showing a Balmer continuum in absorption, whilst the majority of surveyed
SySts present the Balmer continuum in emission, sometimes spectacularly so. 
The SySts showing the Balmer continuum in absorption include V4368 Sgr (a
symbiotic nova at maximum brightness when the burning shell around the WD
expanded to mimic an A-F supergiant, Munari 2019), TX CVn (the A/B spectral
type of the companion is explained by Kenyon and Webbink as originating from
a low-mass white dwarf ($\sim$0.27~M$_\odot$) with a rejuvenated hydrogen
burning shell), and V694~Mon (the A spectral type of the companion comes
from a large accretion disk forming around a massive ($\geq$0.9~M$_\odot$)
white dwarf; Lucy et al.  2020).  We believe that in THA 15-31 the Balmer
continuum in absorption originates in the accretion disk around the
companion, as discussed below in sect.  3.7.

\subsection{Distance}

   \begin{figure}
   \centering
   \includegraphics[width=8.8cm]{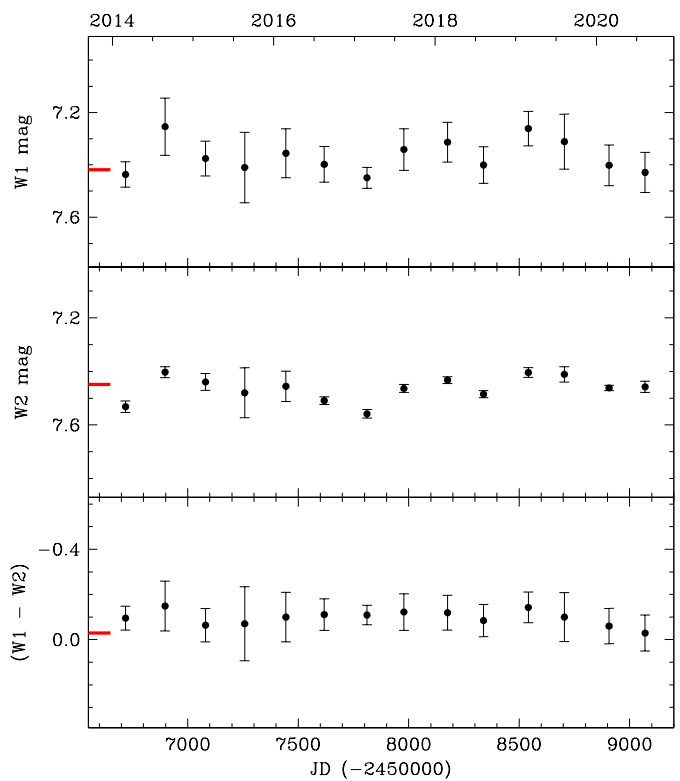}
      \caption{Epoch photometry of THA~15$-$31 in the infrared W1 and W2
      bands as recorded by the NeoWISE mission.  The thick red horizontal mark
      is the corresponding value listed in the AllWISE catalog 
      (see Sect. 3.3 for details).}
         \label{fig:NeoWISE}
   \end{figure}

   \begin{figure}
   \centering
   \includegraphics[width=8.8cm]{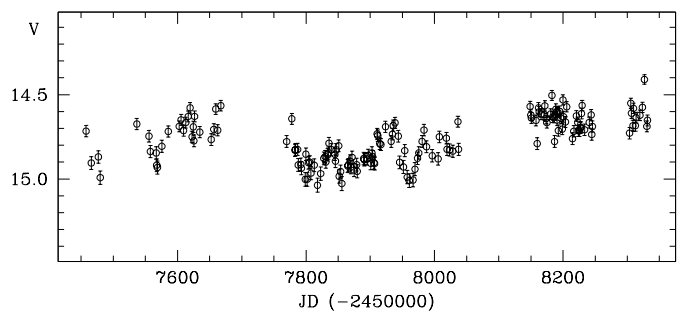}
      \caption{ASASSN $V$-band light curve of the blend composed by THA~15$-$31 
      and the nearby field star 2MASS 16083631$-$4016120, recorded together as a
      single unresolved source.}
         \label{fig:ASASSN}
   \end{figure}

We mention above how the very small parallax given in Gaia EDR3 indicates
a large distance to THA~15$-$31 (at least several kpc).  The
associated error is however too large for a safe inversion of the parallax
to derive an accurate value for the distance (cf.  Bailer-Jones 2015; Luri
et al.  2018).  To estimate the latter, we compare THA~15$-$31 with SU Lyn,
a prototype for accreting-only symbiotic stars that harbors a similar
M6III which suffers from a low $E_{B-V}$=0.05$\pm$0.02 (Mukai et al.  2016,
Lallement et al.  2014, Gontcharov \& Mosenkov 2018).  SU Lyn is nearby and
therefore its Gaia EDR3 parallax is accurate at 1.373$\pm$0.063~mas, which
fixes the absolute 2MASS $K_S$ magnitude of SU Lyn to
M($K_S$)=$-$7.72$\pm$0.11, where the error propagates the uncertainties on
$K_S$, the parallax and the reddening.  Assuming that the M6III giant in
THA~15$-$31 is characterized by the same M($K_S$) as the M6III giant in
SU~Lyn, the distance to THA~15$-$31 turns out to be 12$\pm$0.6~kpc, a value
that we adopt in the remainder of this paper.  The corresponding height
above the galactic plane is 1.8~kpc, which matches the median value for
known SySts believed to be old-population objects (Munari et al.  2021,
their Figure~7).

\subsection{Photometric stability}

THA~15$-$31 does not appear to be significantly variable at optical or
IR wavelengths.
The RG of THA~15$-$31 can be probed with the epoch IR photometry
performed by the Wide-field Infrared Survey Explorer (WISE) over its four
photometric bands W1, W2, W3, and W4 (3.4, 4.6, 12 and 22 $\mu$m,
respectively).  While observations in W3 and W4 bands could be performed
only during the initial cryogenic phase before the depletion of hydrogen
coolant (the first 10 months following launch in 2009), a four-month mission
extension was subsequently conducted in the remaining W1 and W2 bands.  The
AllWISE catalog (Cutri et al.  2013) combines observations from these
2009-2010 cryogenic and post-cryogenic survey phases, while NeoWISE (Mainzer
et al.  2011, 2014) refers to the data the satellite is collecting in W1 and
W2 bands since it was brought out of hibernation and resumed
observations in 2014.  The long-term light curve of THA~15$-$31 in the W1 and
W2 bands is plotted in Figure~\ref{fig:NeoWISE}.  The
IR brightness of THA~15$-$31 appears to have remained relatively constant over the
monitored 2009-2010 and 2014-2020 time intervals.  A Fourier analysis
suggests the presence of a sinusoidal signal of $\sim$0.06 mag
semi-amplitude and $\sim$1400~day period in the W2 data, albeit at low
statistical significance.  However, the currently available NeoWISE data
do not yet cover two full cycles of this possible long periodicity.  The
W1 data are affected by errors larger than the amplitude of such a possible
periodicity, and cannot be used to confirm that it is real rather than an
artefact.

If the RG dominates the flux at longer wavelengths, the relative
contribution of HC grows moving blueward and optical photometry is strongly
influenced by it (cf.  spectral fitting in Figure~\ref{fig:M6III} and the
SED discussed in Sect.  3.4 below).  We therefore inspected public databases
of patrol all-sky surveys in search for epoch optical photometry of
THA~15$-$31.  The object is unfortunately located too far south for the ZTF
survey (Masci et al.  2019, Bellm et al.  2019), which is conducted with the
Palomar Schmidt telescope and has a deep limiting magnitude and a sharp
angular resolution.  On the other hand, the ASASSN survey (Shappee et al. 
2014, Kochanek et al.  2017) covers the whole sky nightly in $V$ or $g$
bands, but the large adopted pixel size (7.8 arcsec) means that THA~15$-$31
cannot be resolved from a nearby field star of comparable magnitude located
11 arcsec away (2MASS~16083631$-$4016120), and the two stars are measured
together as an unresolved pair by ASASSN.  Gaia EDR3 reports $B_P$=15.18 and
16.08 mag for THA~15$-$31 and the nearby star, respectively.  The ASASSN
light curve for this blended source is presented in Figure~\ref{fig:ASASSN}
and is characterized by a limited dispersion in brightness (rms=0.126 mag)
around a mean value ($V$=14.76 mag) that could be somewhat fluctuating over
a long timescale (by $\sim$0.2 mag).  How much of this rather limited
variability is real and intrinsic to THA~15$-$31 is hard to say, and it
could be due to the field star, or instead to the blending of two stellar
images into one.  Certainly, variable seeing and focusing can be expected to
affect the result of the aperture-photometry performed by ASASSN on such an
unresolved but obviously not-single source.

In summary, recorded photometric data suggest that THA~15$-$31 does not
appear to vary significantly, to the point that combining multi-wavelength
data that have been gathered at different epochs into a single SED seems a
reasonable procedure for deriving some basic properties of the system.

\subsection{Spectral energy distribution}

   \begin{figure}
   \centering
   \includegraphics[width=8.8cm]{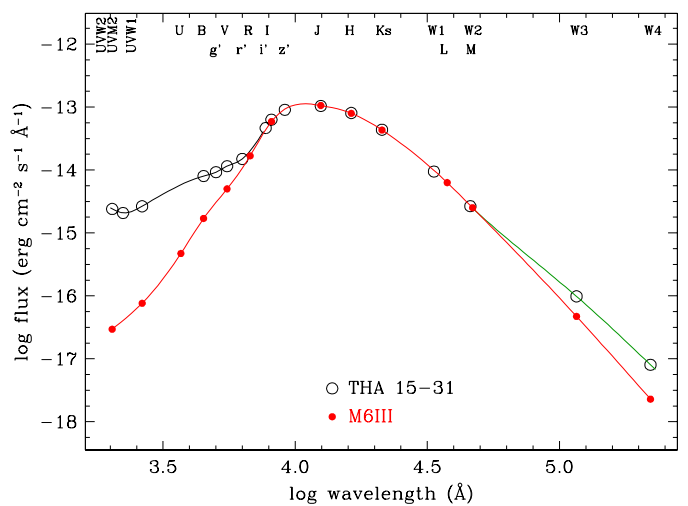}
      \caption{Spectral energy distribution of THA~15$-$31 (black dots and
      black/green line) corrected for $E_{B-V}$=0.38 reddening and compared to that of
      a normal, single M6III star (red dots and line).}
         \label{fig:SED}
   \end{figure}

The reddening-corrected SED of THA~15$-$31 is
presented in Figure~\ref{fig:SED}, where it is compared to that of an M6III
giant built from the intrinsic colors listed by Bessell (1990), Koornneef et
al.  (1983), Lee (1970), and Wu et al.  (1982).  The absence of significant
variability by either the RG or the HC, as discussed in the previous
section, allows us to build the SED of THA~15$-$31 by combining
non-simultaneous data from different sources: AllWISE W1, W2, W3, and W4
magnitudes (Cutri et al.  2013), 2MASS JHKs (Cutri et al.  2003), APASS DR8
BVg'r'i' (Henden \& Munari 2014), SkyMapper z' (Wolf et al.  2018), and the
Swift UVOT UVW1, UVM2, and UVW2 data from our observations as described in
Sect.  2 above.

   \begin{figure}
   \centering
   \includegraphics[width=7cm]{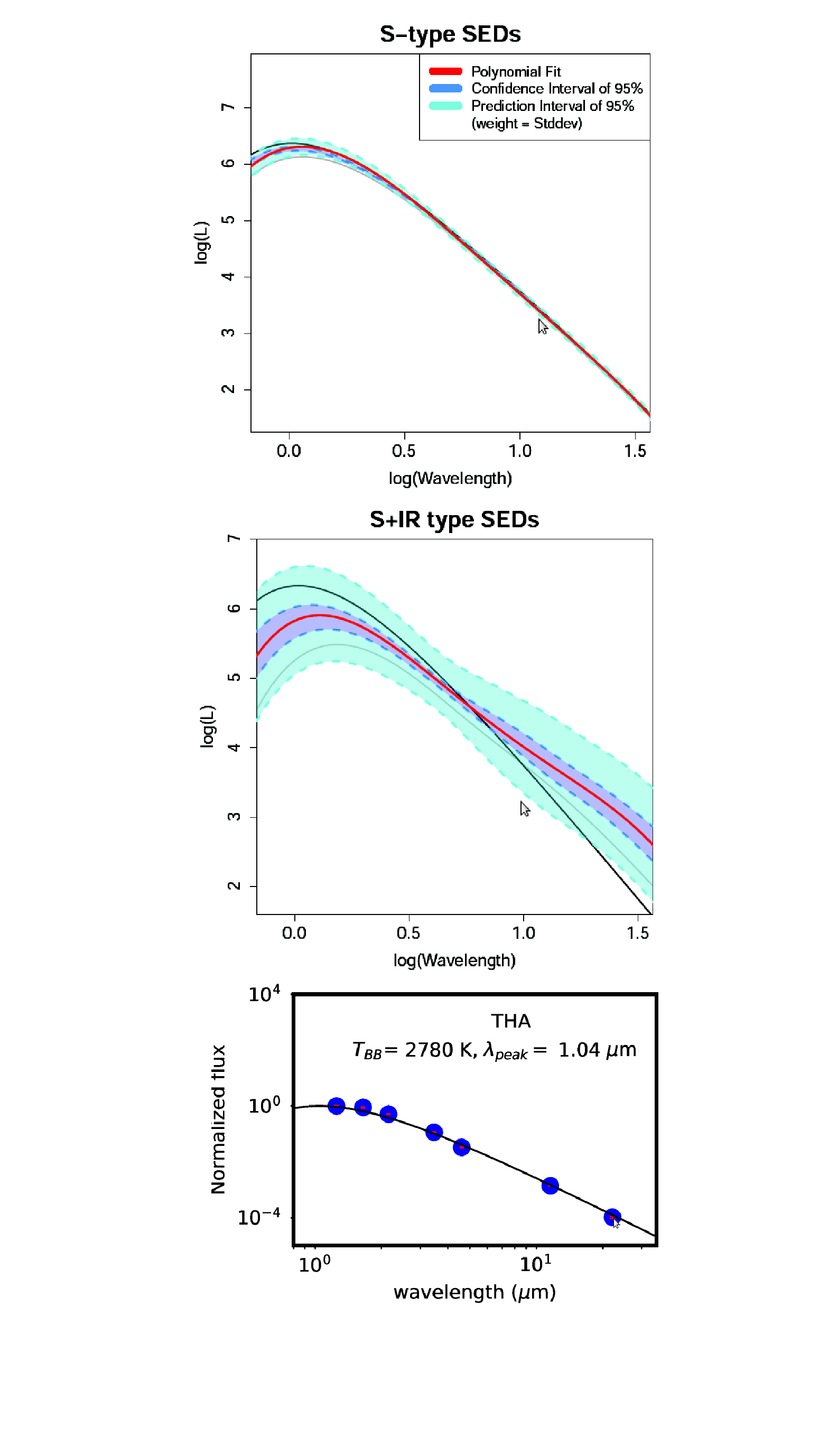}
      \caption{THA 15$-$31 in the near infrared.  {\em Top and central
      panels:} THA 15$-$31 (black line) is overplotted to Figure~4 from Akras
      et al.  (2019a).  {\em Bottom panel:} Energy distribution of THA 15$-$31
      fitted with a single blackbody in the same fashion as Figure~3 of
      Akras et al.  (2019a).  See discussion in sect.  3.4.}
         \label{fig:Akras}
   \end{figure}

The SED in Figure~\ref{fig:SED} shows that THA~15$-$31 matches well a field
M6III in the red and in the NIR, confirming the results from the
spectral fitting in Figure~\ref{fig:M6III}.  Integrating the flux
distribution under the red line in Figure~\ref{fig:SED}  for the
M6III giant in THA~15$-$31 returns a luminosity of
\begin{equation}
L_{\rm M6III} = 6400~L_\odot ,
\end{equation}
corresponding to a bolometric magnitude M$_{bol}$=$-$4.75 which  closely matches
the expectation for a M6III (cf.  Strai{\v{z}}ys 1992, Sowell et al.  2007). 

The contribution by HC to the SED of THA~15$-$31 is rather obvious in
Figure~\ref{fig:SED}, with the observed flux distribution floating well
above that of the RG in the optical and UV.  The difference for
$\lambda \leq$ 7000~\AA\ between the interpolating black and red lines in
Figure~\ref{fig:SED} corresponds to a radiated luminosity of
\begin{equation}
L_{\rm HC} = 100~L_\odot 
.\end{equation} 
A value frequently cited in the literature for other SySts is the $L_{\rm UV}$
luminosity radiated over the three UVOT photometric bands (cf Mukai et al. 
2016): for THA~15$-$31 this value is 6.3$\times$10$^{34}$ erg~s$^{-1}$ or
\begin{equation} 
L_{\rm UV} = 17~L_\odot 
,\end{equation}
which places THA 15-31 among the {\it acc}-SySts accreting at high rates. 
Lopes de Oliveira et al.  (2018) found $L_{\rm UV}$ to vary between 0.3 and
2.8~L$_\odot$ for SU Lyn, and its median value for the accreting SySts
studied by Luna et al.  (2013) is 5~L$_\odot$, while Luna et al.  (2018)
derived $L_{\rm UV}$=25~L$_\odot$ for RT Cru.  Flickering in the UV is
ubiquitous among acc-SySts (Luna et al.  2013), but there is far less
flickering at optical wavelengths (Zamanov et al.  2017), with its detection
requiring the greatest care (Munari et al.  2021).  The acc-SySts with the
largest-amplitude flickering are generally those with the highest $\dot
{M}_{\rm acc}$, and this bodes well for a search in THA 15-31.
 
The SED in Figure~\ref{fig:SED} shows an excess emission over the WISE W3
and W4 bands, which is probably caused by dust.  Integrating the difference
between the red and green lines in Figure~\ref{fig:SED} for $\lambda \geq$
5~$\mu$m allows us to derive the luminosity of this IR excess as
\begin{equation}
L_{\rm dust} = 35~L_\odot 
.\end{equation} 
Such dust does not add to the extinction of interstellar origin affecting
THA~15$-$31 as defined in Sect.~3.1 above, meaning it either does not
project over the M6III star or, if it engulfs the central binary, it is
located at a sufficiently large distance to dilute the dust grains below the
column density required to impact the observed reddening.

An energy distribution fitted by a stellar photosphere over the entire 1$-$4
$\mu$m range (bands $J$ through $L$) is labeled $S$-type following the
terminology introduced by Allen \& Glass (1974), as opposed to $D$-type when
emission from dust at 800-1000~K dominates instead.  The majority of known
SySts are of the $S$-type, while SySts of the $D$-type usually harbor an RG
affected by Mira pulsating variability.  Later, Allen (1982) introduced a
third class, the $D^\prime$-type, where emission from dust at 400-500~K
causes an excess in $L$-band (3.5 $\mu$m) while $J$$H$$K$ bands remain
dominated by the stellar photosphere.  

A fourth class, termed {\it S+IR}, was recently introduced by Akras et al. 
(2019a, hereafter Ak19) and is characterized by a {large} excess emission
detectable only over the 12$-$22 $\mu$m range covered by the WISE W3 and W4
bands.  Ak19 compiled an updated catalog of known SySts and classified them
into the four types: $S$, {\it S+IR}, $D^\prime$, and $D$.  Their criteria
would have grouped THA 31-15 with the $S$-type SySts, as illustrated in
Figure~\ref{fig:Akras} where our program star is overplotted on the original
Ak19 Figure~4 and its IR energy distribution is fitted in the same fashion
as their Figure~3.  Nonetheless, THA 15-31 presents a clear excess emission
in W3 and W4 bands, and an association with the pure $S$-type objects would
under-represent this.  To classify SySts in IR, Ak19 fitted their energy
distributions with simple blackbodies: one for the stellar photosphere,
another for the dust.  The latter was only required when the excess caused
by dust was too large to be accommodated within a stretching of the fit with
just the stellar-photosphere blackbody.  Thus, in the Ak19 reference frame,
the transition from pure $S$ to {\it S+IR} type occurs only for really large
excess emission at W3 and W4 bands: a moderate excess as displayed by THA
15-31 is not distinguished from a pure stellar photosphere.  THA 15-31 is
the first identified example of a symbiotic star with a moderate amount of
excess in W3 and W4 bands, a {transition object} between SySts of really
pure $S$-type and the {\it S+IR} ones.  It would be fair to expect that
other SySts similar to THA 15-31 exist among those grouped into the $S$-type
by Ak19.

We conclude this section by noting that the X-ray nondetection of THA
15-31 is easily explained by its inferred distance.  Indeed, the
$\sim$1500~s integration we performed with Swift/XRT, which implies an
X-ray luminosity upper limit of $<$1.1$\times$10$^{33}$
($<$1.4$\times$10$^{33}$ unabsorbed) in the 0.3--10 keV band, would not have
led to the X-ray detection of SU~Lyn (the proto-type of the low-$L_{\rm
acc}$ SySts; Mukai et al.  2016) or EG~And (representing the high-$L_{\rm
acc}$ SySts; Nu\~nez et al.  2016) if they were at the distance of
THA~15$-$31, which is approximately 20 times more distant than the two above
examples, thus reducing the observed flux by a factor of 400 and
bringing it below the sensitivity threshold of XRT.

\subsection{Emission lines}

   \begin{figure*}
   \centering
   \includegraphics[angle=270,width=17.5cm]{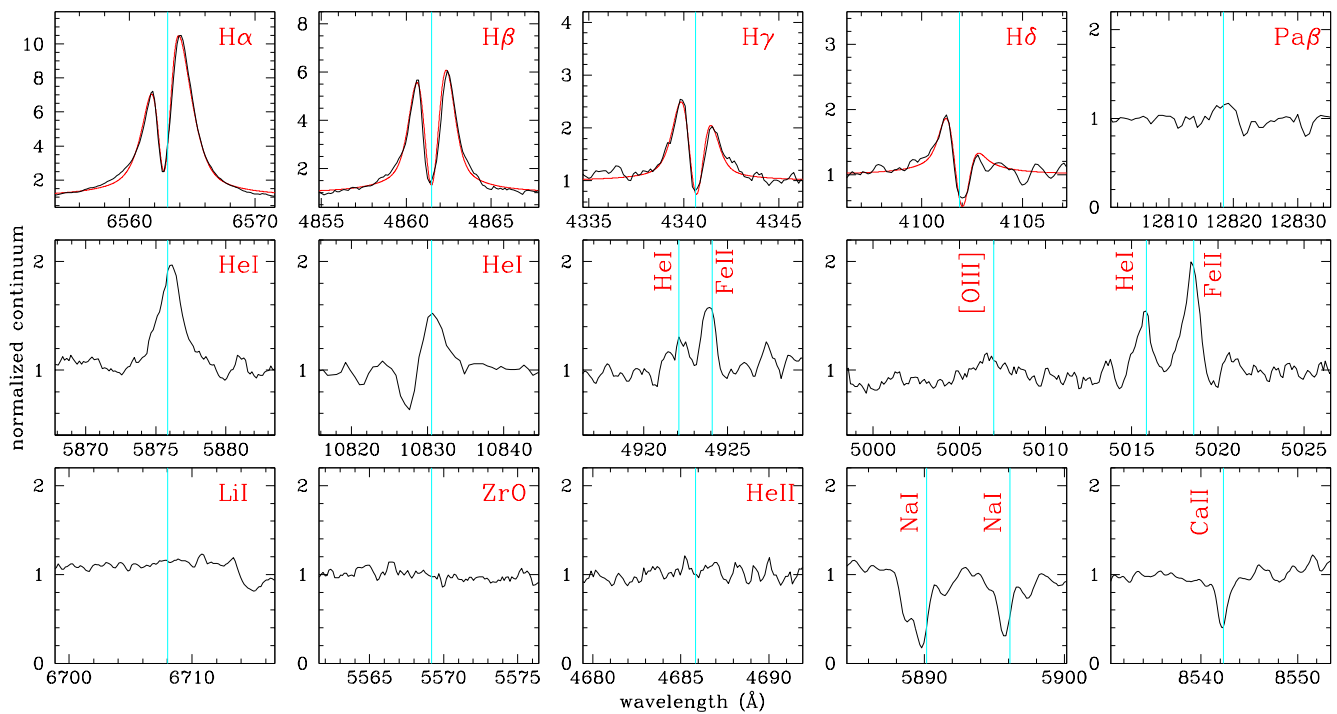}
      \caption{Line profiles for THA~15$-$31.  {\it Top and central rows:}
      Profiles for a sample of the emission lines recorded on the X-Shooter
      spectrum of THA~15$-$31.  {\it Bottom row:} Some additional spectral
      features from the same spectrum and discussed in the text.  All panels
      are plotted over the same velocity scale.  The thin vertical lines
      mark the laboratory wavelength in the rest frame of the red giant.}
         \label{fig:line_profiles}
   \end{figure*}

A sample of the recorded emission line profiles of THA 15-31 is presented in
Figure~\ref{fig:line_profiles}: Balmer lines show broad wings and a sharp absorption
superimposed on the emission component, while all other lines are well
fitted by a single Gaussian in emission.  The P-Cyg profile for HeI
10830 is only apparent and is not genuine: the absorption to the left is a normal
photospheric line from the M6III star (SiI 10827 \AA; cf.  Vaughan \& Zirin
1968).  Table~\ref{em_lines} lists the radial velocity and integrated flux of
observed emission lines.

The profile of Balmer lines was de-convolved with a Voigt profile for the
emission and a Gaussian for the absorption, and the resulting fit is
over-plotted on the observed profile in Figure~\ref{fig:line_profiles}.  The
principal parameters of the fitting are listed in Table~\ref{balmer}.  Given
the weakness of the emission component of H$\gamma$ and H$\delta$, the
corresponding values in Table~\ref{balmer} are much more uncertain than for
H$\alpha$ and H$\beta$.  There is an obvious progression moving up along the
Balmer series: the absorption component is on the blue side of the emission
in H$\alpha$ and progressively turns to the red side for higher terms.  This
behavior is not generally observed in SySts, for which the difference in
velocity between emission and absorption components usually remains the same
for different Balmer lines.  However, we sometimes observe a pattern in
SySts that is similar to that displayed by THA 15-31; for example as
displayed by CH Cyg and Hen 2-468 on the spectra we collected on July 24,
2016, with the Asiago 1.82m+Echelle telescope.  We do not have a simple
explanation at hand for such behavior.  In most SySts, the absorption
component is seen to follow the orbital motion of the RG with an offset of
$-$10/$-$20 km~s$^{-1}$, suggesting it forms in the wind that blows off the
RG and engulfs the whole binary.  In any case, the radial velocities
reported in Table~\ref{em_lines} for the emission lines of THA 15-31 other
than Balmer suggest that they form in a stratified medium: the mean velocity
for HeI singlets is +7.5$\pm$0.5 and for triplets is +20.5$\pm$1
km~s$^{-1}$; it is $-$3$\pm$1 km~s$^{-1}$ for FeII multiplet \#42, and
+12.5$\pm$1~km~s$^{-1}$ for multiplet \#49.

\begin{table}[t]
\small
\begin{center}
\caption{Heliocentric wavelength of line photo-center, corresponding
heliocentric radial velocity, and integrated flux for emission lines
identified in the X-Shooter spectrum of THA 15-31.  Pa-$\alpha$ is overly
affected by saturated telluric absorptions, preventing a more meaningful measurement.}
\label{em_lines}
\begin{tabular}{lccrc}
\hline\hline
line &      & $\lambda\odot$ & RV$_\odot$ & flux    \\
     &      & (\AA)     &(km s$^{-1}$) & (erg cm$^{-2}$ s$^{-1}$)  \\
\hline
HeI  & sing & 4922.068  &  8.5  & 1.33e-15    \\ 
HeI  &  "   & 5015.770  &  5.7  & 1.42e-15    \\
HeI  &  "   & 6678.328  &  8.0  & 1.81e-15    \\
HeI  &  "   & 20581.80  &  7.6  & 1.54e-14    \\
HeI  & trip & 5876.053  & 20.5  & 5.53e-15    \\
HeI  &  "   & 7065.802  & 22.3  & 8.75e-15    \\
HeI  &  "   & 10830.85  & 18.8  & 9.58e-14    \\
FeII & \#42 & 4923.877  &$-$2.7 & 1.67e-15    \\
FeII &  "   & 5018.407  &$-$1.6 & 2.36e-15    \\
FeII &  "   & 5168.946  &$-$4.9 & 2.02e-15    \\
FeII & \#49 & 5234.800  & 10.3  & 5.74e-16    \\
FeII &  "   & 5276.258  & 15.0  & 5.46e-16    \\
FeII &  "   & 5316.825  & 12.2  & 2.11e-15    \\
H$\delta$  && 4101.289  &$-$32.8& 1.17e-15    \\
H$\gamma$  && 4340.415  &$-$3.7 & 5.84e-15    \\
H$\beta$   && 4861.565  & 14.4  & 3.12e-14    \\
H$\alpha$  && 6563.379  & 25.7  & 3.38e-13    \\
Pa-$\gamma$&& 10938.65  & 15.2  & 9.17e-15    \\
Pa-$\beta $&& 12818.63  & 12.9  & 3.17e-14    \\
Pa-$\alpha$&& present   &       &$>$1.2e-13   \\
\hline
\end{tabular} 
\end{center}
\end{table} 

\begin{table}[t]
\small
\begin{center}
\caption{Parameters of the fitting to the Balmer lines in Figure~5 with 
one Gaussian for the absorption and one Voigt profile for the emission. For
the latter, the FWHM of the Gaussian core and of the Lorentzian wings are
listed.}
\label{balmer}
\begin{tabular}{@{}c@{~~}c@{~~}c@{~~}c@{~~}c@{~~}c@{~~}c@{}}
\hline\hline
line         & \multicolumn{3}{c}{emission} &
             & \multicolumn{2}{c}{absorption} \\ \cline{2-4} \cline{6-7}
             & RV$_\odot$   
             & FWHM  
             & FWHM &     
             & RV$_\odot$    
             & FWHM  \\
             &      & core& wings     &&       &       \\ 
\hline
H$\alpha$    & 18   & 145 &  225      && $-$1  &  54   \\ 
H$\beta$     & 12   & 105 &  160      &&   12  &  71   \\ 
H$\gamma$    &  1:  & 112:&  150      &&   11  &  65   \\ 
H$\delta$    &  4:  &  86:&  145      &&   13  &  69   \\ 
\hline
\end{tabular} 
\end{center}
\end{table}

The lack of essential information for this object, such as orbital period,
orbital inclination, or the presence of flickering, prevents us from
speculating as to where and how the emission lines form in THA 15-31, if in
the accretion disk, in the ionized wind of the M6III, or both.  If they form
primarily in the main body of an accretion disk, the single-Gaussian shape
of the nonBalmer lines argues in favor of a disk seen at low inclination
(face-on).  A high inclination disk (edge-on) could be accommodated if, at
the time of X-Shooter observation, the view to a significant fraction of the
disk was obstructed (as, e.g., during ingress or egress from an eclipse
behind the M6III giant).  On the other hand, the presence of a
fast-developing flickering could argue in favor of the presence of a
luminous hot spot at the outer rim of the disk, and this could be a relevant
site for emission-line formation (as is inferred from the so called "S-wave"
orbital modulation observed in cataclysmic variables; e.g., Hellier 2001). 
The presence of a hot spot would in turn imply that the M6III likely fills
its Roche lobe, and mass transfer proceeds primarily via overflow at L1 with
the resulting stream impacting the disk at the location of the hot spot. 
The higher and harder the luminosity radiated by the HC, the larger the
fraction of the RG wind it can ionize; in turn this would shift the primary
source of emission-line formation from the disk toward the ionized wind. 
The orbital modulation of the radial velocity and of the line profiles would
be different in the two cases, and long-term, high-resolution spectroscopic
monitoring would be clearly beneficial in identifying the location of
formation of the emission lines and in constraining other basic parameters
of THA 15-31.

As a final comment, we note that the H$\alpha$ and H$\beta$ emission lines
in Figure~\ref{fig:line_profiles} would have well satisfied the selection
criteria in Munari et al.  (2021) if THA 15-31 were in GALAH, in addition to
the spectral signatures of an M-giant and the absence of radial pulsations
in the light curve.  The profile of THA 15-31 H$\alpha$ is a fine match to
that of the 33 GALAH SySts, and the condition on equivalent widths
EW(H$\alpha$)$>$EW(H$\beta$) is met within a large margin.  The EW of
H$\alpha$ and H$\beta$ is one order of magnitude larger in THA 15-31 than in
the 33 GALAH SySts, and the luminosity radiated by the respective accreting
stars scales by the same proportion.

\subsection{Other spectral features}

The bottom row of Figure~\ref{fig:line_profiles} shows the X-Shooter
spectrum at some interesting wavelengths.  The CaII line is well centered at
the M6III radial velocity, with a clean profile.  Considering that CaII and
FeII have a similar ionization potential, this suggests that FeII emission
lines do not form in or near the M6III atmosphere.  On the other hand, the
NaI doublet is instead blueshifted (by $-$9.7$\pm$0.6 and $-$10.4$\pm$0.6
km~s$^{-1}$ for D1 and D2 components, respectively).  Considering that (1)
the equivalent width of the interstellar component corresponding to
$E_{B-V}$=0.38 reddening is rather large at 0.53~\AA\ (from the calibration
by Munari and Zwitter 1997), and that (2) the general radial velocity of the
interstellar medium is expected to be negative along the line of sight to
THA 15-31 (Brand \& Blitz 1993), the observed $-$10~km~s$^{-1}$ radial
velocity for NaI could result from the unresolved blend between interstellar
and stellar components.  The HeII panel in Figure~\ref{fig:line_profiles}
clearly shows the absence of this line, the hallmark of high-ionization
conditions in symbiotic stars.  The ZrO panel is centered at the head of the
deepest absorption band from this molecule, which is usually the strongest
among $s$-element oxides in optical spectra (Keenan 1954); its absence
suggests the M6III giant has not (yet) undergone the third dredge-up and has
not been polluted in $s$-elements by the progenitor of the current WD
companion (cf.  Jorissen 2003 for extrinsic $s$-element stars among SySts). 
Finally, no LiI is present in the spectrum of THA 15-31.

\begin{table}[t]
\small
\begin{center}
\caption{Summary of the basic parameters of THA~15$-$31.}
\label{summary}
\begin{tabular}{rll|rl}
\hline\hline
      &alt.name &Sz105       &long  (Gal)  &338.724       \\
RA    &(J2000)&16:08:36.90   &lat   (Gal)  &+08.516       \\
DEC   &(J2000)&$-$40:16:20.4 &        dist&12   kpc      \\
APASS &B      &16.35         &  $z$-height&1.8  kpc      \\
      &V      &15.06         &   $E_{B-V}$&0.38          \\
Gaia  &G      &12.85         &spec. type&M6\,III \\
      &B$_P$  &15.17         &  RV$_\odot$(M6III)&4 km\,s$^{-1}$\\
      &R$_P$  &11.46         &L$_{\rm M6III}$&6400  L$_\odot$\\
2MASS &J      &9.02          &L$_{\rm HC}$&$\geq$100  L$_\odot$\\
      &H      &8.01          &L$_{\rm dust}$&$\geq$35  L$_\odot$\\
      &K$_S$  &7.59          &    M(V$_S$)&$-$1.65 \\
UVOT  &UVW1   &17.86         &    M(K$_S$)&$-$7.65 \\
      &UVM2   &19.10         &  types&acc-only\\
      &UVW2   &18.86         &       &dusty\\
\hline
\end{tabular} 
\end{center}
\end{table} 

\subsection{The source of UV excess}

Although quite relevant, the $L_{\rm HC}$=100~L$_\odot$ luminosity estimated
above for the hot component of THA 15-31 is too low to be ascribed to stable
nuclear burning on the surface of a WD, a condition believed to hold for a
significant fraction of known SySts (Kenyon 1986, Sokoloski 2003).  The
conditions for stable and steady H-burning at the surface of an accreting WD
have been explored, among others, by Nomoto et al.  (2007), Shen \& Bildsten
(2007), and Wolf et al.  (2013).  In these models, the lowest burning
luminosities, $L_{burn}$=1740~L$_\odot$, pertain to the lightest WD models
of 0.51~M$_\odot$ accreting at 2.5$\times$10$^{-8}$~M$_\odot$\,yr$^{-1}$,
while for a WD of 1.34~M$_\odot$ accreting at
1.1$\times$10$^{-6}$~M$_\odot$\,yr$^{-1}$ the burning luminosity reaches the
upper limit of $L_{burn}$=81,000~L$_\odot$.  A similar argument would hold
against residual shell-burning from a previous nova outburst, a scenario
also contrasting with the stable brightness seen when comparing the
present-day THA 15-31 to the Palomar DSS-1 plates of the 1950s.

Alternative to nuclear burning on a WD, the UV excess observed in THA 15-31
may originate from ionized circumstellar material; a normal stellar
companion; or an accretion disk.  We consider them in turn, and conclude in
favor of the latter.

The inset of Figure~\ref{fig:M6III} shows a Balmer continuum in marked
{absorption}.  Ionized gas glowing under recombination would inevitably
present the Balmer continuum in emission (Osterbrock \& Ferland 2006), not
in absorption: this suffices to exclude nebular material ---such as an
ionized fraction of the wind from the M6III--- as the primary source of the
UV excess in THA 15-31.

Several reasons also argue against a normal star as the source of UV excess. 
The emission lines of HeI observed in THA 15-31 originate from upper levels
$\sim$23 eV above the ground state.  This and the lack of HeII emission
lines suggests excitation temperatures in the range
2.5$-$5.0$\times$10$^{4}$~K.  For normal stars, such a range corresponds to O
and early-B spectral types, for which the radiated luminosity (10$^{3}$ to
10$^{4}$~$L_\odot$; Drilling \& Landolt 2000) appears too large in
comparison to the 100~L$_\odot$ of the HC in THA 15-31.  Another argument
against a normal star is the Balmer series visible in absorption up to H(29)
in the inset of Figure~\ref{fig:M6III}.  It is well known (e.g.,  Inglis \&
Teller 1939) that the quantum number of the last term observed in the Balmer
series relates to the electronic density; in practical terms (cf.  Jaschek
\& Jaschek 1987; Munari et al.  2005), the last Balmer line observable is
$\sim$H(16) for hot main sequence stars and $\sim$H(19) for their
high-luminosity counterparts. H(18) or H(19) are also the last term
visible in emission in nonbinary Mira variables (A.  Siviero, priv. 
comm.).  The H(29) recorded in THA 15-31 suggests an electronic density
much lower than encountered in the stellar atmospheres of normal stars, and
even more so for the high-surface-gravity OB subdwarfs.

Further evidence against the hypothesis that THA 15-31 is a normal star is
the energy distribution of the hot component in Figure~\ref{fig:SED}, which
cannot be approximated by any stellar model for O- or B-type stars or by a
black body at a given temperature: the observed shape resembles instead a
gray-body distribution, of the type that may originate from the combined
emission of the anuli of an accretion disk, each emitting at a different
temperature (cf.  La Dous 1989).  Finally, binaries composed of cool giants
and hot normal star companions are named VV~Cep variables, from the
prototype (Pantaleoni Gonz{\'a}lez et al.  2020); as discussed in Munari et
al.  (2021) such objects are rather massive and very young, with all known
members of the class laying close to the plane of the Galaxy, with a median
height of just 50pc.  The much greater height of THA 15-31, 1.8 kpc from the
plane, is incompatible with the young age of the O- and B-type stars there.

We are then left with an accretion disk as the source for the UV excess
observed in THA 15-31.  In support of this, we note that on the Asiago
Echelle spectra that we regularly obtain on V694 Mon, the highest Balmer line
visible in absorption is either H(28) or H(29), similarly to THA 15-31. 
V694 Mon is a well-known symbiotic star that attracted much attention in
1990 when, for some months, it displayed short-living and distinct broad
absorption lines distributed in velocity up to 6000 km/s (Tomov 1990). 
There is a general consensus that V694 harbors an accreting WD, with its
massive disk dominating at optical and UV wavelengths (Shore et al.  1994,
Zamanov et al.  2011, Lucy etal.  2020), and confirmed by its large
flickering activity (Michalitsianos et al.  1993, Tomov et al.  1996) and
the characteristic hard X-ray emission from the boundary layer (Stute \& Sahai
2009).

   \begin{figure}
   \centering
   \includegraphics[width=8.8cm]{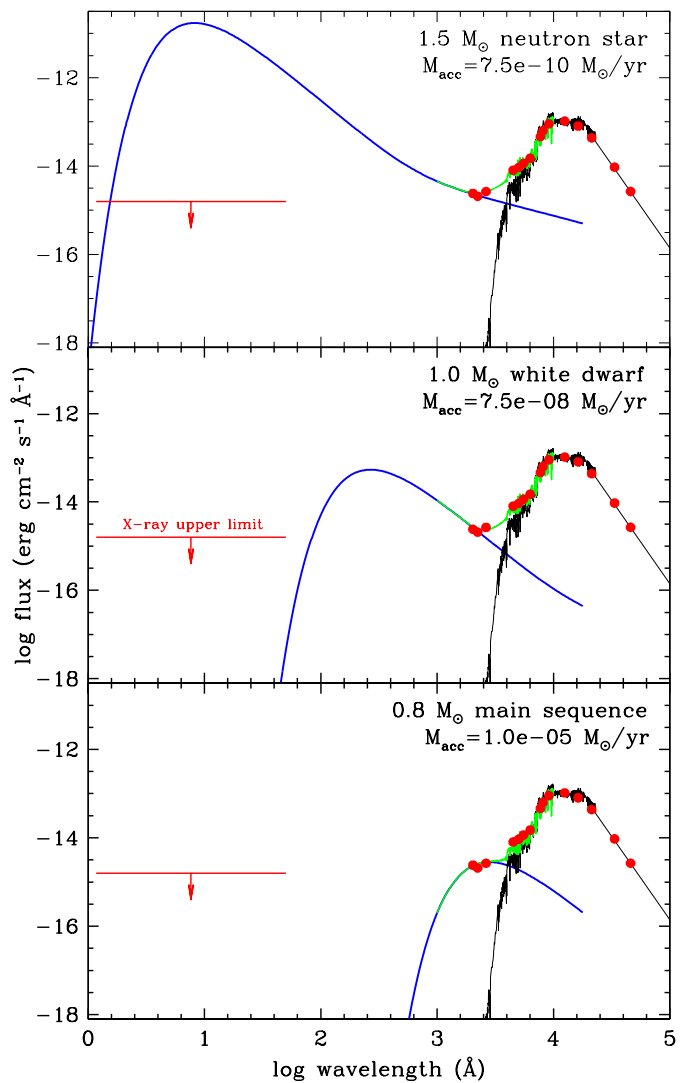}
        \caption{Examples of fitting the SED of THA
        15-31 combining the M6III with an accretion disk (and its central
        object) around a neutron star (top), a white dwarf (center), and a
        main sequence star (bottom).  Details are discussed in Sect.  3.8. 
        The arrow on the left marks the upper limit in X-rays from Swift-XRT
        observations.  The reference distribution for the M6III is the same
        as that used in Figure~\ref{fig:M6III}.}
         \label{fig:disk}
   \end{figure}

\subsection{The accreting star}

Having established an accretion disk as the likely source of the UV excess
emission, the next step is to constrain the nature of the central object. 
We lack crucial observations at wavelengths intermediate between Swift UVOT
and XRT intervals, which would be of high diagnostic value here. 
Nonetheless, available data allow us set useful constraints.

We computed accretion disk models to fit the UV excess observed in THA
15-31 for different choices of the central object, always of the
nonmagnetic variety: a neutron star, a white dwarf, and a main sequence
star.  The models are adapted from the implementation presented in
Mucciarelli et al.  (2007) and Patruno \& Zampieri (2008), with the disk
that extends from the surface of the central object up to twice the
circularization radius, and includes self-irradiation.  Examples of the
resulting SED are presented in Figure~\ref{fig:disk}.  

The computed luminosity and energy distribution depend on a number of
parameters, including: the masses of the central and the donor stars, the
binary period, the accretion rate, the SED of the donor, in addition to the
inclination angle and the orbital phase of the binary system.  For THA
15-31, most of them are unknown, and reasonable ranges have been extensively
explored.  The results presented in Figure~\ref{fig:disk} are typical of
each class, and can therefore be regarded as guidelines for these three
configurations: their energy distribution invariably peaks in the X-rays for
a neutron star, in the extreme UV for a white dwarf, and in the near-UV for
a main sequence star.

At the top of Figure~\ref{fig:disk}, we consider the case of a neutron star
of 1.5~M$_\odot$  and accreting at 7.5e-10~M$_\odot$/yr.  Whatever the mass
of the neutron star or the accretion rate, fitting the flux in the Swift
UVOT bands {always} produces an X-ray emission far in excess of the upper
limit observed in THA 15-31 with Swift XRT.  Introducing a magnetic field
for the neutron star would only exacerbate the problem, channeling the inner
parts of the disk onto the polar caps and increasing the X-ray flux with the
emission from the accretion columns.  Therefore, we can confidently exclude
a neutron star as the accreting object in THA 15-31.

At the bottom of Figure~\ref{fig:disk}, we explore a main sequence (MS) star
as the central object, with a mass of 0.8~M$_\odot$ and accreting at
1e-05~M$_\odot$/yr.  As already noted by Kenyon and Webbink (1984), workable
MS models for SySts invariably require furious mass-accretion rates, and our
simulations certainly confirm this.  The very shallow potential well of an
MS star, whilst easily accounting for the absence of any X-ray emission,
requires very high $\dot{M}_{\rm acc}$ to produce enough flux to fit THA
15-31 in the UV.  The temperature attained by the main body of the disk is
generally too low to support photoionization at energies larger than for
hydrogen.  Harder photons can be produced at the boundary layer, but the
high optical thickness of such a disk suggests that they are all absorbed
locally with few escaping beyond the boundary layer itself.  The modeling of
the boundary layer under the very large accreting rates required by an MS is
rather uncertain, and a simple scaled-up version of the formalism generally
adopted for WD or NS could be misleading.  For such a reason, we have not
included a boundary layer in the disk model for the MS case at the bottom of
Figure~\ref{fig:disk}.

At the center of Figure~\ref{fig:disk}, we finally consider a WD as the
accreting object.  In the plotted case, the WD has a mass of 1.0~M$_\odot$
and accretes at 7.5e-08~M$_\odot$/yr.  The WD is slowly rotating, has a
radius of 0.008 $R_\odot$, and the accretion efficiency and the albedo of
the disk and donor surface layers are fixed at $2.6 \times 10^{-4}$ and 0.9,
respectively.  The mass adopted for the M6III is 1.2~M$_\odot$, 1000~days
for the orbital period, and $i=60^\circ$ for the orbital inclination.  With
such a configuration, the Roche lobe of the donor is comparable to the
radius of the M6III giant ($\sim 200$-$300$ R$_\odot$).  The bolometric
luminosity of the disk is $150 \, L_\odot$, consistent with the estimate
reported in equation (3).  The emission from the disk well accounts for the
Swift/UVOT flux and peaks in the extreme UV, providing a significant
reservoir of ionizing photons for the formation of the hydrogen emission
lines (there are also enough $\lambda$$\leq$228~\AA\ photons to expect
detection of HeII 4686.  The spectra of THA 15-31 do not show this line, but
they were taken several years before the acquisition of the Swift
observations modeled here, and changes in the accretion rate between the
two epochs may explain the difference).  At the same time, the spectrum cuts
off well below the soft X-rays, which is consistent with the lack of
detection with Swift/XRT.  We did not perform a fit because it would
not be particularly constraining for the accretion disk (considering that
most of the SED is dominated by the red supergiant).  However, varying
${\dot M}$, $i,$ and the albedo, we verified that the observed fluxes can be
satisfactorily reproduced in a range of values close to those reported
above, which shows that the result is relatively robust.  In particular, a
value of $\dot{M}_{\rm acc}$=2e-08~M$_\odot$ is also
acceptable for a face-on system ($i\sim 0^\circ$) with an albedo of 0.8.  We note
that the accretion rate is above Eddington, but the bolometric disk
luminosity is not because the efficiency is low.

The mass-loss rate of isolated, nonpulsating M giants is generally assumed
to be a few 10$^{-7}$ M$_\odot$\,yr$^{-1}$ (e.g., Livio \& Warner 1984). 
The above requirements on $\dot{M}_{\rm acc}$ for the case of an accreting
WD in THA 15-31 would be easily met by conventional Roche-lobe overflow or
by the {wind} Roche-lobe overflow suggested by the hydrodynamical
simulations of Mohamed \& Podsiadlowski (2012).  If the orbital separation
in THA 15-31 is too large for the RG to approach its Roche-lobe, the
accretion needs to proceed via wind-intercept.  Assuming the wind leaves the
RG and travels distances longer than the orbital separation in a spherically
symmetric arrangement, the standard Bondi-Hoyle treatment for wind-intercept
(Livio \& Warner 1984) shows that the companion captures only a few percent
of the material lost by the RG, too low a fraction to account for the high
$L_{\rm acc}$ observed in THA 15-31.  This fraction needs to be one order of
magnitude higher than allowed by the Bondi-Hoyle scenario, a condition
frequently encountered in SySts (e.g., Skopal 2005).  This can be explained
by the companion's gravitational focusing of the wind leaving the RG toward
the orbital plane (e.g., Decin et al.  2019).  The latter results in a
strong density enhancement on the orbital plane at the expense of the polar
directions: the higher density favors a higher accretion by the companion
compared to the case of an unfocussed, spherically symmetric wind outflowing
from the RG.  The presence of a strong density enhancement on the equatorial
plane of SySts is expected theoretically (Mohamed \& Podsiadlowski 2007,
Booth et al.  2016), and was confirmed by the marked bipolar shape observed
for the spatially resolved nova ejecta of RS~Oph (Ribeiro et al.  2009) and
V407 Cyg (Giroletti et al.  2020), and inferred by modeling of line profiles
in high-inclination systems like EG~And by Shagatova et al.  (2016, 2021).

In conclusion, an accretion disk around a WD provides the simplest and
easiest explanation for the UV excess seen in THA 15-31, with a large space
of parameters able to reproduce observations, and with the required
mass-loss rate from the cool giant being a fine match to normal values. 
While there is no room for a NS in THA 15-31, a disk around an MS star could
still fit the UVOT observations but only while invoking an accretion rate on
the MS star of at least two orders of magnitude {larger} than the mass-loss
rate typical of nonpulsating M giants.  A rate of 1e-05~M$_\odot$/yr is
achieved only at the tip of the AGB, where it may be supported by radial
pulsations (H{\"o}fner \& Olofsson 2018).  These evolved giants are usually
enshrouded in massive circumstellar cocoons abundant in dust, and may show
over-abundance in $s$-element following third dredge-up episodes.  Such an
AGB is unlikely present in THA 15-31: no radial pulsations, or large dust
excess, or even $s$-element overabundance is observed, and as noted in
Figure~\ref{fig:M6III} the spectral match with normal M6III giants is
instead perfect.  Therefore, while the presence of an MS star in THA 15-31
cannot be excluded with absolute certainty, it would require that such an
odd series of conditions be met that it is (highly) unlikely.

The brightness of THA 15-31 on the blue plates of the first Palomar all-sky
photographic atlas was similar to current values, suggesting it was
accreting to a comparable rate during the 1950s.  It would be interesting to
reconstruct its long-term light curve, for example from inspection of archive
photographic plates, and ascertain for how long the current $\dot{M}_{\rm
acc}$ has been maintained.  The presence or absence of past, unnoticed
outbursts could then be compared with the stability of accretion disks in
SySts over time, and the amount of material piled-up on the companion with
the amount necessary to ignite nuclear burning on the surface of a WD.

Finally, the previous misclassification of  THA 15-31  as a young stellar
object suggests that other accreting-only symbiotic stars may be lurking in
old YSO catalogs.  Such misclassification is not unprecedented: a
significant fraction of symbiotic stars of the burning type were originally
discovered and classified as bona fide planetary nebulae: it was only the
discovery at later times and on IR wavelengths of the presence of a cool
giant in the system that led to rectification of their initial
classification (e.g., Allen 1983, Acker \& Stenholm 1990).  In regard to
this, we find that the IR selection criteria for SySts, derived by Akras et
al.  (2019b) following a machine learning approach, nicely distinguish THA
15-31 from true YSOs in the list of H$\alpha$ emission stars for the Lupus
and Scorpius regions compiled by The (1962).

\begin{acknowledgements}
We would like to thank the anonymous Referee for constructive
suggestions that led to improve the paper.  A.~Skopal provided useful
comments on an early version of this paper, and A.~Siviero let us to access
unpublished data from his atlas of high resolution spectra of Mira
variables. We thank Jamie Kennea and the Swift team for the quick approval
and the rapid acquisition of the observations we requested.  NM acknowledges
financial support through ASI-INAF and 'Mainstream' agreement 2017-14-H.0
(PI: T.  Belloni).  JMA and AF acknowledge financial support from the
project PRIN-INAF 2019 Spectroscopically Tracing the Disk Dispersal
Evolution (STRADE).  This work was supported by the Swedish strategic
research programme eSSENCE.  GT was supported by the project grant ‘The New
Milky Way’ from the Knut and Alice Wallenberg Foundation and by the grant
2016- 03412 from the Swedish Research Council.  GT also acknowledges
financial support of the Slovenian Research Agency (research core funding
no.  P1-0188 and project N1-0040).  This publication makes use of data
products from the Near-Earth Object Wide-field Infrared Survey Explorer
(NEOWISE), which is a joint project of the Jet Propulsion
Laboratory/California Institute of Technology and the University of Arizona. 
NEOWISE is funded by the National Aeronautics and Space Administration.
\end{acknowledgements}

\end{document}